\documentclass[]{eptcs}

\usepackage{breakurl}             
\usepackage{underscore}           

\usepackage{mathptmx}
\usepackage{amsmath}
\usepackage{color}
\usepackage{amssymb}
\usepackage{graphicx}
\usepackage[T1]{fontenc}
\usepackage{xspace}
\usepackage{cite}
\usepackage{amsthm}

\def\prem{$\cal P$$\!reM$\xspace}

\def\iprem{\emph{i}\prem}

\def\rprem{\emph{r}\prem}

\newcommand{\bldl}{\smallskip\[\begin{array}{ll}}
\newcommand{\cldl}{\[\begin{array}{ll}}
\newcommand{\eldl}{\end{array}\]\rm}

\def\magg#1{min \langle #1 \rangle}
\def\amagg#1{cMin \langle #1 \rangle}
\def\mahh#1{max \langle #1 \rangle}
\def\amahh#1{cMax \langle #1 \rangle}
\def\cntv#1{count \langle #1 \rangle}

\newcommand{\arrow}{~\mbox{\tt <-}~}

\title{BigData Applications from Graph Analytics to Machine Learning by Aggregates in Recursion}
\author{Ariyam Das \quad Youfu Li \quad Jin Wang \quad Mingda Li \quad Carlo Zaniolo
\institute{Department of Computer Science}
\institute{University of California, Los Angeles, USA}
\email{\{ariyam, youfuli, jinwang, limingda, zaniolo\}@cs.ucla.edu}
}
\def\titlerunning{Complex Analytics by Aggregates in Recursion}
\def\authorrunning{Das et al.}
\begin{document}
\maketitle
\begin{abstract}
    

In the past, the  semantic issues raised by the  non-monotonic nature of aggregates often prevented  their use in the recursive statements
 of logic programs and deductive databases.  However, the recently introduced notion of Pre-mappability (\prem) has shown that, in key applications of interest, aggregates can be used in recursion to optimize  the perfect-model semantics of  aggregate-stratified programs. Therefore we can preserve the declarative formal semantics of such programs while achieving a highly efficient operational semantics that  is conducive to scalable implementations on parallel and distributed platforms. In this paper, we show that with \prem, a wide spectrum of classical algorithms of practical interest, ranging from graph analytics and dynamic programming based optimization problems to data mining and machine learning applications can be concisely expressed in declarative languages by using aggregates in recursion. 
Our examples are also used to show that \prem can be checked using simple techniques and templatized verification strategies. 
A wide range of advanced BigData applications can now be expressed declaratively in logic-based languages, including  Datalog, Prolog, and even SQL, while enabling their execution with superior performance and scalability \cite{rasql},\cite{ssp-iclp19}.

\end{abstract}
\section{Introduction}\label{intro}
Prolog's  success with advanced applications  demonstrated the ability of declarative languages to express powerful algorithms as  ``logic + control.''
Then, after observing that in relational database management systems, ``control'' and optimization are provided by the system implicitly, Datalog  researchers sought the ability to express powerful applications using only declarative logic-based constructs.  After initial successes, which e.g., led to the introduction of recursive queries in SQL, Datalog  encountered two major obstacles  as data analytics grew increasingly complex:
  (i) lack of expressive power at the language level, and 
  (ii) lack of scalability and performance at the system level.
 
These problems became clear with the rise of more complex descriptive and  predictive BigData analytics. For instance, the in-depth study of data mining algorithms   \cite{db2CacheMining} carried out in the late 90s by the IBM DB2 team concluded that the best way to carry out predictive analytics is to load the data from an external database into main memory and then write an efficient implementation in a procedural language to mine the data from the cache.
However, recent advances in architectures supporting in-memory parallel and distributed computing have led to the renaissance of powerful declarative-language based systems like \emph{LogicBlox} \cite{logicblox}, \emph{BigDatalog} \cite{bigdatalog}, \emph{SociaLite} \cite{socialite}, \emph{BigDatalog-MC} \cite{bigdatalog-mc}, \emph{Myria} \cite{Myria} and \emph{RASQL} \cite{rasql} that can scale efficiently on multi-core machines as well as on distributed clusters.
In fact, some of these general-purpose systems like \emph{BigDatalog} 
and \emph{RASQL} have outperformed commercial graph engines like \emph{GraphX} for many classical graph analytic tasks.
%
This has brought the focus back on to the first challenge (i) -- how to express the wide spectrum of predictive and prescriptive analytics in declarative query languages. This problem has assumed great significance today with the revolution of machine learning driven data analytics, since ``in-database analytics'' can save data scientists considerable time and effort, which is otherwise repeatedly spent in extracting features from  databases via multiple joins, aggregations and projections and then exporting the dataset for use in external learning tools to generate the desired analytics \cite{relationalAI}.   
Modern researchers have worked toward this ``in-database analytics'' solution by writing \emph{user-defined functions} in procedural languages or using other low-level system interfaces, which the query engines can then import \cite{bismarck}. However this approach raises three fundamental challenges: 

\begin{itemize}
    \item \textbf{Productivity and Developability:} Writing efficient implementations of advanced data analytic applications (or even modifying them) using low-level system APIs require data science knowledge as well as system engineering skills. This can strongly hinder the productivity of data scientists and thus the development
of these advanced applications. 
    \item \textbf{Portability:} User-defined functions written in one system-level API may not be directly portable to other systems where the architecture and underlying optimizations differ.  
    \item \textbf{Optimization:} Here, the application developer is entrusted with the responsibility to write an optimal user-defined function, which is contrary to the work and vision of the database community in the 90s \cite{Imielinski} that aspired for a high-level declarative language like SQL supported by implicit query optimization techniques. 
\end{itemize}

In this paper, we argue that these problems can be addressed by simple extensions that enable the use of aggregate functions in the recursive  definitions of  logic-based languages, such as Datalog, Prolog, and even SQL.  To that effect, we use different case studies to show that simple aggregates in declarative recursive computation can  express concisely and declaratively a host of advanced applications ranging from graph analytics and dynamic programming (DP) based optimization problems to data mining and machine learning (ML) algorithms. While the use of non-monotonic aggregates in recursive programs raises difficult semantic issues, the newly introduced notion of \emph{pre-mappability} (\prem) \cite{zaniolo-amw2016} can ensure the equivalence of former programs with that of aggregate-stratified programs under certain conditions. Following this notion of \prem, we further illustrate step-by-step how a data scientist or an application developer can very easily verify the semantic correctness of the declarative programs, which provide these complex ML/AI-powered data analytic solutions.
Before diving into these case studies, let us briefly introduce \prem.
\vspace{-0.2cm}\section{Pre-Mappable Constraints in Graph Queries}\label{background}
 We consider a Datalog query, given by rules $r_{1.1}-r_{1.3}$, to compute the shortest paths between all vertices in a graph given by the relation 
 {\tt arc(X, Y, D)}, where $D$ is the distance between vertices $X$ and $Y$. In this query, as shown in rule $r_{1.3}$, the aggregate 
 {\tt min} is defined on group-by variables $X$ and $Y$, at a stratum higher than the recursive rules ($r_{1.1}$ and $r_{1.2}$).
 Thus, we use the compact head notation often used in the literature
for aggregates.
 %
\begin{displaymath}
\begin{aligned}
& r_{1.1}: {\tt path(X, Y, D)} \arrow {\tt arc(X, Y, D)}. \\
& r_{1.2}: {\tt path(X, Y, D)} \arrow {\tt path(X, Z, Dxz)}, {\tt arc(Z, Y, Dzy)},  {\tt D=Dxz+Dzy}. \\
& r_{1.3}: {\tt shortestpath(X, Y, \magg{D})} \arrow {\tt path(X, Y, D)}. \\
\end{aligned}
\end{displaymath}
%
The \texttt{min} and \texttt{max} aggregates can also be viewed as 
constraints enforced upon the results returned in the head of the rule:
i.e., for the example at hand the \emph{min} constraint $\tt (X, Y, \magg{D})$
is enforced on $\tt shortestpath(X, Y, D)$. This view allows us to
define the semantics of  $r_{1.3}$ by re-expressing it with negation as shown in rules $r_{1.4}$ and $r_{1.5}$. This guarantees that the program has a perfect-model semantics, although the iterated fixpoint computation of such model  can be very inefficient and even non-terminating in presence of cycles.
\begin{displaymath}
\begin{aligned}
& r_{1.4}: {\tt shortestpath(X, Y, D)} \arrow {\tt path(X, Y, D)}, {\tt \neg betterpath(X, Y, D)}. \\
& r_{1.5}: {\tt betterpath(X, Y, D)} \arrow {\tt path(X, Y, D)}, {\tt path(X, Y, Dxy)}, {\tt Dxy < D}.
\end{aligned}
\end{displaymath}
The aforementioned inefficiency can be cured with \prem, whereby the {\tt min} aggregate can be pushed inside the recursion within the same stratum, as shown in rules $r_{2.1}$ and $r_{2.2}$. Because of  \prem this transformation is  equivalence-preserving  \cite{scalingup-tplp2018} since the program below has a minimal fixpoint and computes the $\tt shortestpath$ atoms of the original program in a finite number of iterations. 

In general, this transformation holds true for any constraint $\gamma$ and Immediate Consequence Operator (defined over recursive rules) $T$, if $\gamma(T(I)) = \gamma(T(\gamma(I)))$, for every interpretation $I$ of the program.
\begin{displaymath}
\begin{aligned}
& r_{2.1}: {\tt path(X, Y, \magg{D})} \arrow {\tt arc(X, Y, D)}. \\
& r_{2.2}: {\tt path(X, Y, \magg{D})} \arrow {\tt path(X, Z, Dxz)}, {\tt arc(Z, Y, Dzy)},  {\tt D=Dxz+Dzy}. \\
& r_{2.3}: {\tt shortestpath(X, Y, D)} \arrow {\tt path(X, Y, D)}. \\
\end{aligned}
\end{displaymath}
Testing that  \prem was satisfied during the execution of a program is straightforward \cite{rasql}. 
Furthermore, simple formal tools \cite{zaniolo-amw2018} are at hand to prove that \prem holds for any possible execution of a given program, but due to space limitations we will simply use the reasoning embedded in those tools to prove \prem for the cases at hand. 
For example, \prem is always  satisfied by base rules such as $r_{2.1}$  \cite{zaniolo-tplp2017}, and hence we only need to prove the property for the recursive rule $r_{2.2}$,  i.e.
we prove that the additional constraint $\Bar{\gamma} = {\tt (X, Z,\magg{Dxz}}$)
can be imposed on $I ={\tt path(X, Z, Dxz)}$ without changing the result returned in 
the head in as much as this is constrained by $\gamma = {\tt (X, Y,\magg{ Dxz+Dzy}}$).
Indeed  every ${\tt (X, Z, Dxz)}$ that violates $\Bar{\gamma}$  produces a ${\tt ( X, Y, D)}$ value that violates 
$\gamma$ and it is thus eliminated. So the addition of $\Bar{\gamma}$ does not change the result when 
$\gamma$  is also in place. An even more dramatic situation occurs, if we replace 
$\tt D=Dxz+Dzy$ with, say, $\tt D = 3.14 * Dzy$   in our recursive rule.  Then it is clear that
the result computed in the head of the rule is invariant w.r.t the value of $\tt Dxz$,
and therefore we could even select the \texttt{min} of these values. 
In other words, we here have  that  $T(I)= T(\gamma(I))$.  Obviously this is a special case of \prem, that will be called {\em intrinsic \prem} (or \iprem in short). 
Another special case of \prem, called {\em radical \prem} (or \rprem in short) occurs when the equality 
 $\gamma(T(I)) = T(\gamma(I))$ holds. This is for instance the case when the  condition $\tt X=a$ is added to the rule $r_{2.3}$, which specifies that we are only  interested in the paths that originate in $\tt a$. Then this condition can be
 pushed all the way to the non-recursive base rule $r_{2.1}$, leaving the recursive rule unchanged and thus amenable to the \texttt{min} optimization previously described. While the use of \rprem in pushing constants was widely studied in the Datalog literature, the use of \iprem  and full \prem in dealing with non-monotonic constraints was introduced in \cite{zaniolo-tplp2017}.
%
%
\section{Dynamic Programming based Optimization Problem}\label{dpo}
Consider the classic \emph{coin change} problem: given a value \texttt{V} and an infinite supply of each of $C_1, C_2, ..., C_n$ valued coins, what is the minimum number of coins needed to get change for \texttt{V} amount?
Traditionally, declarative programming languages attempt to solve this through a stratified program: the lower stratum recursively enumerates over all the possible ways to make up the value \texttt{V}, while the \texttt{min} aggregate is applied at the next stratum to select the desired answer. 
Obviously, such simple stratified recursive solutions are computationally extremely inefficient.
In procedural languages, these problems are solved efficiently with dynamic programming (DP) based optimization. Such DP based solutions utilize the ``optimal substructure property'' of the problem i.e., the optimal solution of the given problem can be evaluated from the optimal solutions of its sub-problems, which are, in turn, progressively calculated and stored in memory (memoization). For example, consider an extensional predicate \texttt{coins} having the atoms \texttt{coins(2)}, \texttt{coins(3)} and \texttt{coins(6)}, which represent coins with values 2 cents, 3 cents and 6 cents respectively. Now, we need at least 2 coins to make up the value $V = 9$ cents (3 cents + 6 cents). Note, we can also make up 6 cents using 3 coins of 2 cents each. However, the optimal solution to make up 9 cents should also in turn use the best alternative available to make up 6 cents, which is to use 1 coin of 6 cent itself. 
Based on this discussion, the example program below, described by rules $r_{3.1}-r_{3.2}$, shows how this solution can be succinctly expressed in Datalog with aggregate in recursion. This program can be executed in a top-down fashion and 
the optimal number of coins required to make up the change is determined by passing the value of \texttt{V} (9 in our example) to the recursive predicate \texttt{num} (as shown by the query goal). 
\begin{displaymath}
\begin{aligned}
& r_{3.1}: {\tt num(C, 1)} \leftarrow {\tt  coins(C)}.\\
& r_{3.2}: {\tt num(V, \magg{N})} \leftarrow {\tt  coins(C)}, {\tt  C < V}, {\tt X = V - C}, {\tt num(X, Y)}, {\tt N = Y + 1}.\\
& ?-{\tt num(9, N)}.
\end{aligned}
\end{displaymath}

The successive bindings for the predicate \texttt{num} are calculated from the coin value \texttt{C} under consideration (as \texttt{V - C}) and are passed in a top-down manner (top-down information passing) till the exit rule $r_{3.1}$ is reached. The \texttt{min} aggregate inside recursion ensures that for every top-down recursive call (sub-problem) only the optimal solution is retained. With this said materialization of the intensional predicate \texttt{num} (analogous to memoization), this program execution is almost akin to a DP based solution except one difference --- pure DP based implementations are usually executed in a bottom-up manner.
%
In the same vein, it is worth mentioning that many interesting DP algorithms (e.g., computing minimum number of operations required for a chain matrix multiplication) can also be effectively computed with queries, containing aggregates in recursion, using bottom-up semi-naive evaluation identical to the DP implementations. 
We next focus our attention on validating \prem for the above program. Note the definition of \prem, \iprem or \rprem does not refer to any evaluation strategy for processing the recursive query i.e. the definitions are agnostic of top-down, bottom-up or magic sets based recursive query evaluation strategies.
Interestingly, the use of ``optimal substructure property'' in DP algorithms itself guarantees the validity of \prem. 
This can be illustrated as follows with respect to the \texttt{min} constraint: consider inserting an additional constraint 
$\Bar{\gamma} = {\tt num(X, \magg{Y}}$)
on $I ={\tt num(V, N)}$ in the recursive rule $r_{3.2}$. 
Naturally, any \texttt{Y}, which does not satisfy $\Bar{\gamma}$, will produce a \texttt{N} that violates the \texttt{min} aggregate in the head of rule $r_{3.2}$ and hence will be discarded. Since, the imposition of $\Bar{\gamma}$ in the rule body does not change the result when 
$\gamma$ in the head (of rule $r_{3.2}$) is applied, the \texttt{min} constraint can be pushed inside recursion i.e., $\gamma(T(I)) = \gamma(T(\gamma(I)))$, thus validating \prem.  
%

\section{K-Nearest Neighbors Classifier}\label{knn}


$K$-nearest neighbors is a popular non-parametric instance-based lazy classifier, which stores all instances of the training data. Classification of a test point is computed based on a simple majority vote among $K$ nearest\footnote{Based on metrics like Euclidean distance.} training instances of the test point, where the latter is assigned into the class that majority of the $K$ neighbors belong to.

In the Datalog program, defined by rules $r_{4.1}-r_{4.7}$, the predicate \texttt{te(Id,X,Y)} denotes a relational instance of two-dimensional test points  represented by their \texttt{Id} and coordinates \texttt{(X,Y)}. Likewise, the predicate \texttt{tr(Id,X,Y,Label)} denotes the relational instance of training points 
represented by their \texttt{Id}, coordinates \texttt{(X,Y)} and corresponding class \texttt{Label}. In this example, rule $r_{4.1}$ calculates the Euclidean distance between the test and all the training points, while the recursive rule $r_{4.3}$ with aggregate determines the nearest $K$ neighbors for each of the test point. Symbolically, the predicate \texttt{nearestK(IdA,D,IdB,J)} represents the training instance \texttt{IdB} is the \texttt{J}-th nearest neighbor of the test point \texttt{IdA} located at a distance of \texttt{D} apart. Finally, rules $r_{4.4}-r_{4.5}$ aggregates the votes for different classes and performs the classification by majority voting. \texttt{cMax} in rule $r_{4.5}$ is a special construct that extracts the corresponding class \texttt{Label} that received the maximum votes for a given test point.  Rule $r_{4.5}$ can be alternatively expressed without \texttt{cMax}, as shown in rules $r'_{4.5}, r''_{4.5}$. 
%
In terms of simple relational algebra, the constructs \texttt{cMin} or \texttt{cMax} can be thought of denoting the projection of specific columns (attributes like ${\tt Id_2}$ in $r_{4.3}$ and ${\tt Label}$ in $r_{4.5}$) from a tuple, which satisfies the \texttt{min} or \texttt{max} aggregate constraint respectively. However, these special constructs are mere syntactic sugar as illustrated before with equivalent rules 
$r'_{4.5}, r''_{4.5}$, which do not use any of these constructs. 
%
\begin{displaymath}
\begin{aligned}
& r_{4.1}: {\tt dist(Id_1, Id_2, D)} \leftarrow {\tt  te(Id_1, X_1, Y_1)}, 
{\tt  tr(Id_2, X_2, Y_2, Label)}, {\tt  D = (X_1-X_2)^2 + (Y_1-Y_2)^2}. \\
& r_{4.2}: {\tt nearestK(Id, -1, -1, nil)} \leftarrow {\tt te(Id, X, Y)}. \\
& r_{4.3}: {\tt nearestK(Id_1, \magg{D}, \amagg{Id_2}, J_1)} \leftarrow 
{\tt dist(Id_1, Id_2, D)}, {\tt nearestK(Id_1, S, Id_3, J)}, \\
& \hspace{200pt} {\tt larger(S, Id_3, D, Id_2)}, {\tt  J_1 = J+1}, {\tt J_1 \leq K}. \\
& r_{4.4}: {\tt votes(Id_1, Label, \cntv{Id_2})} \leftarrow 
{\tt nearestK(Id_1, D, Id_2, J)}, 
{\tt tr(Id_2, X, Y, Label)}. \\
& r_{4.5}: {\tt classify(Id_1, \mahh{V}, \amahh{Label})} \leftarrow 
{\tt votes(Id_1, Label, V)}. \\
& r_{4.6}: {\tt larger(S, Id_3, D, Id_2)} \leftarrow 
{\tt D > S}. \\
& r_{4.7}: {\tt larger(S, Id_3, D, Id_2)} \leftarrow 
{\tt D = S, Id_2 > Id_3}.
\end{aligned}
\end{displaymath}
%
%
\begin{displaymath}
\begin{aligned}
& r'_{4.5}: {\tt classify(Id_1, V, Label)} \leftarrow 
{\tt votes(Id_1, Label, V)}, \neg {\tt higher(Id_1, V)}. \\
& r''_{4.5}: {\tt higher(Id_1, V)} \leftarrow 
{\tt votes(Id_1, Label, V)}, {\tt votes(Id_1, Label', W)}, {\tt W > V}. 
\end{aligned}
\end{displaymath}

We now verify that the \texttt{min} aggregate in the recursive rule $r_{4.2}-r_{4.3}$ satisfies \prem  and ensures semantic correctness.
Note the exit rule $r_{4.2}$ always trivially satisfies the \prem definition, since the interpretation, $I$ of the recursive predicate is initially an empty set. Thus, we focus our attention only on the recursive rule $r_{4.3}$. We now prove that $r_{4.3}$ satisfies \iprem: consider inserting an additional constraint $\tt{(Id_1, J, \magg{S})}$ in the body of the rule $r_{4.3}$ that defines the \texttt{min} constraint on the recursive predicate \texttt{nearestK} in the body (creating an interpretation $\gamma(I)$ in the rule body). If this \texttt{min} constraint in the body ensures that for a given $\tt{Id_1}$ and $\tt{J}$, $\tt{S}$ is the minimum distance of the $\tt{J}$-th nearest neighbor, then for the corresponding valid $\tt{J_1} (\leq K)$, $r_{4.3}$ without the \texttt{min} aggregate in the head will produce all potential $\tt{J_1}$-th neighbors whose distances are higher than $\tt{S}$ (i.e., distance of $\tt{J}$-th neighbor), thereby being identical to $T(I)$. Thus, we have, $T(I) = T(\gamma(I))$ validating $r_{4.3}$ satisfies \iprem, since the recursive rule remains invariant to the inclusion of the additional constraint $\tt{(Id_1, J, \magg{S})}$ in the rule body.  
%
%

Similar to $K$-nearest neighbor classifier, several other data mining algorithms like $K$-spanning tree based graph clustering, vertex and edge based clustering, tree approximation of Bayesian networks, etc. --- all depend on the discovery of a sub-sequence of elements in sorted order and can likewise  be expressed with \prem using aggregates in recursion. It is also worth observing that while our declarative $K$-nearest algorithm requires more lines of code than the other cases presented in this paper, it can still be expressed with only seven lines of logical rules as compared to standard learning tools like \emph{Scikit-learn} that implements this in 150+ lines of  procedural or object-oriented code.
%
%
\section{Iterative-Convergent Machine Learning Models}\label{mlmodel}
Iterative-convergent machine learning (ML) models like SVM, perceptron, linear regression, logistic regression models, etc. are often trained with batch  gradient descent and can be written declaratively as Datalog programs with XY-stratification, as shown in \cite{xyStratificationML}. 
Rules $r_{5.1}-r_{5.3}$ show a simple XY-stratified program template to train a typical iterative-convergent machine learning model. 
\texttt{J} denotes the temporal argument, while \texttt{training\_data} (in $r_{5.2}$) is an extensional predicate representing the training set and 
\texttt{model(J, M)} is an intensional predicate defining the model 
\texttt{M} learned at iteration \texttt{J}. The model is initialized using the predicate \texttt{init\_model} and the $X$-rule $r_{5.2}$ computes the corresponding error ${\tt E}$ and gradient ${\tt G}$ at every iteration based on the current model and the training data using the predicate \texttt{compute} (defined according to the learning algorithm under consideration). The final $Y$-rule $r_{5.3}$ assigns the new model for the next iteration based on the current model and the associated gradient using the \texttt{update} predicate (also defined according to the learning algorithm at hand). Since many iterative-convergent ML models are formulated as convex optimization problems, the error gradually reduces over iterations and the model converges when the error reduces below a threshold $\delta$.
%
%
\begin{displaymath}
\begin{aligned}
& r_{5.1}: {\tt model(0, M)} \leftarrow {\tt  init\_model(M)}. \\
& r_{5.2}: {\tt stats(J, E, G)} \leftarrow {\tt model(J, M)}, {\tt training\_data(Id, R)}, {\tt compute(M, R, E, G)}. \\
& r_{5.3}: {\tt model(J+1, M')} \leftarrow 
{\tt stats(J, E, G)}, {\tt model(J, M)}, {\tt update(M, G, M')}, 
{\tt E > \delta}.
\end{aligned}
\end{displaymath}
%
Interestingly, an equivalent version of the above program can be expressed with aggregates and pre-mappable constraints in recursion, as shown with rules $r'_{5.1}-r'_{5.4}$. 
The stopping criterion ${\tt \gamma: E > \delta}$ pushed inside the recursion in rule $r'_{5.3}$ satisfies \rprem, 
since $T(\gamma(I))$ and $\gamma(T(I))$ would both generate the same atoms in \texttt{find}, where the error \texttt{E} is above the threshold $\delta$ (assuming convex optimization function). Also note, the \texttt{max} aggregate defined over the recursive predicate \texttt{find} trivially satisfies \iprem. 
%
%
\begin{displaymath}
\begin{aligned}
& r'_{5.1}: {\tt model(0, M)} \leftarrow {\tt  init\_model(M)}. \\
& r'_{5.2}: {\tt stats(J, E, G)} \leftarrow {\tt model(J, M)}, {\tt training\_data(Id, R)}, {\tt compute(M, R, E, G)}. \\
& r'_{5.3}: {\tt find(\mahh{J}, \amahh{M}, \amahh{E}, \amahh{G})} \leftarrow 
{\tt model(J, M)}, {\tt stats(J, E, G)}, {\tt E > \delta}. \\
& r'_{5.4}: {\tt model(J_1, M')} \leftarrow 
{\tt find(J, M, E, G)}, {\tt update(M, G, M')}, {\tt J_1 = J + 1}.
\end{aligned}
\end{displaymath}
\vspace{-0.2cm}\section{Conclusion}\label{conclusion}
Today BigData applications are often developed and operated in silos, which only support a particular family of tasks -- e.g. only descriptive analytics or only graph analytics or only some ML models and so on. This lack of a unifying model  makes development extremely ad hoc, and hard to port efficiently over multiple platforms. For instance, on many graph applications native Scala with Apache Spark cannot match the performance of systems like RaSQL, which can plan the best data partitioning/swapping strategy for the whole query and optimize the semi-naive evaluation accordingly \cite{rasql}.
However, as demonstrated in this paper, a simple extension to declarative programming model, which allows use of aggregates and easily verifiable pre-mappable constraints in recursion, can enable developers to write concise declarative programs (in Datalog, Prolog or SQL) and express a plethora of applications ranging from graph analytics to data mining and machine learning algorithms. This will also increase the productivity of developers and data scientists, since they can work only on the logical aspect of the program without being concerned about the underlying physical optimizations. 
\bibliographystyle{eptcs}
\bibliography{references}
\end{document}